\documentclass[twocolumn,showpacs,preprintnumbers,amsmath,amssymb]{revtex4}

\usepackage{amssymb}
\usepackage{epsfig, verbatim}
\usepackage{graphicx}
\usepackage[figuresright]{rotating}
\usepackage{bm}
\usepackage{amsmath}

\begin{document}

\title{Atomic and photonic entanglement concentration via photonic Faraday rotation}

\author{Zhao-Hui Peng,$^{1,2,}$ Jian Zou,$^{3}$ Xiao-Juan Liu,$^{2}$ Yong-Jun Xiao,$^{1}$ and Le-Man Kuang$^{1,}$\footnote{Corresponding author: lmkuang@hunnu.edu.cn}}

\affiliation{$^{1}$Key Laboratory of Low-Dimensional Quantum
Structures and Quantum Control of Ministry of Education, and
Department of Physics, Hunan Normal University, Changsha 410081, P.
R. China\\
$^{2}$School of Physics, Hunan University of Science
and Technology, Xiangtan 411201, P. R. China\\
$^{3}$School of Physics, Beijing Institute of Technology, Beijing 100081, P. R. China}

\begin{abstract}
We propose two alternative entanglement concentration protocols
(ECPs) using the Faraday rotation of photonic polarization. Through
the single-photon input-output process in cavity QED, it is shown
that the maximally entangled atomic (photonic) state can be
extracted from two partially entangled states. The distinct feature
of our protocols is that we can concentrate both atomic and photonic
entangled states via photonic Faraday rotation, and thus they may be
universal and useful for entanglement concentration in the
experiment. Furthermore, as photonic Faraday rotation works in low-Q
cavities and only involves virtual excitation of atoms, our ECPs are
insensitive to both cavity decay and atomic spontaneous emission.
\end{abstract}


\maketitle

Entanglement is the key resource in quantum information processing
(QIP), such as quantum teleportation \cite{1}, quantum key
distribution \cite{2} and quantum dense coding \cite{3}. In order to
complete such QIP protocols perfectly, the maximally entangled
states are usually required. However, the entanglement will
inevitably degrade in the process of distribution and storage due to
the interaction between system and its external environment. To
overcome the dissipation and decoherence, Bennett \emph{et al.}
proposed the protocols of entanglement purification \cite{4} and
entanglement concentration \cite{5}. By use of entanglement
purification protocols (EPPs), one can distill a set of mixed
entangled states into a subset of highly entangled states with local
operation and classical communication \cite{4}. However, EPPs can
only improve the quality of the mixed state and can not get the
maximally entangled state. On the other hand, entanglement
concentration protocols (ECPs) \cite{5} can be used to convert the
partially entangled pairs to the maximally entangled ones. In the
early days, many efforts have been devoted to photonic ECPs with
linear \cite{6,7} or nonlinear \cite{8} optical elements. Recently,
ECPs of solid state qubits (such as atomic \cite{9,10,11} or
electric qubits \cite{12}) have also been investigated frequently.

Cavity quantum electrodynamics (QED) system \cite{13} is an
excellent platform for understanding the fundamental principle of
quantum mechanics and investigating QIP. In most of QIP protocols
based on cavity QED, they usually require that atoms strongly
interact with high-Q cavity field, which guarantees not only
entanglement preparation but also further implementation of QIP
tasks. However, as the high-Q cavity is well isolated from the
environment, it seems unsuitable for efficiently accomplishing the
input-output process of photons, which is the key step to implement
long-distance QIP in a scalable fashion. Recently, An \emph{et al.}
\cite{14} proposed a novel scheme to implement QIP with a single
photon by an input-output process with respect to low-Q cavities. It
is shown that the different polarized photon can gain different
phase shift when it interacts with the atom trapped in the low-Q
cavity, which is known as Faraday rotation \cite{15}. Due to the
fact that photonic Faraday rotation works in low-Q cavities and only
involves virtual excitation of atoms, it is insensitive to both
cavity decay and atomic spontaneous emission. Following this scheme,
various works including entanglement generation \cite{16}, quantum
logic gate \cite{17} and quantum teleportation \cite{18} have been
presented. To our best knowledge, ECPs in cavity QED mainly focused
on atomic entanglement concentration \cite{9,10,11}, but there is no
report on concentration of photonic entanglement. The main reason is
that in most cases we are only interested in high-Q cavities not the
low-Q ones, and in some cases the cavity mode is even adiabatically
eliminated and thus has no contribution to the system evolution in
the case of large detuning between the cavity field and atoms
\cite{9,19}.

Inspired by Ref. \cite{14}, we investigate ECPs using the Faraday
rotation of photonic polarization. The low-Q cavity and
single-photon pulse (three-level atom) are introduced to assist
concentration of atomic (photonic) entangled state. Through the
single-photon input-output process in cavity QED, we can extract the
maximally entangled atomic (photonic) state from two partially
entangled states. The distinct feature of our proposals is that we
can concentrate both atomic and photonic entangled states via
photonic Faraday rotation, and thus they may be universal and useful
in the experiment. Furthermore, as our ECPs work in low-Q cavities
and only involve virtual excitation of atoms, they are insensitive
to both cavity decay and atomic spontaneous emission, and may be
feasible with current technology.

\begin{figure}[t]
\begin{center}
\scalebox{0.23}[0.22]{\includegraphics{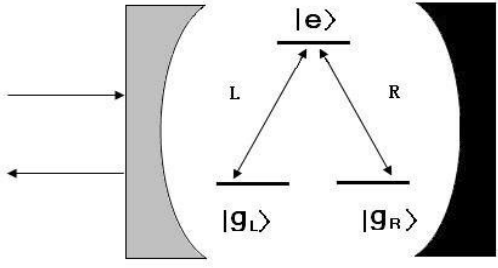}}
\end{center}
\caption{The interaction between three-level atom and a
single-photon pulse propagating input-output the low-Q cavity.}
\end{figure}

Firstly, we briefly review photonic Faraday rotation. Consider a
three-level atom interacting with a low-Q cavity (one-side) driving
by an input photon pulse, as shown in Fig. 1. The atom has two
degenerate ground states ($|g_{L}\rangle$ and $|g_{R}\rangle$) and
an excited state ($|e\rangle$). The transitions
$|g_{L}\rangle{\leftrightarrow}|e\rangle$ and
$|g_{R}\rangle{\leftrightarrow}|e\rangle$ for the atom are assisted
respectively by left-circularly ($L$) and right-circularly ($R$)
polarized photons and the transition frequency is $\omega_{0}$. We
consider the low-Q cavity limit and the weak excitation limit, then
can solve the Langevin equations of motion for cavity and atomic
lowering operators analytically. Adiabatically eliminating the
cavity mode, we obtain the reflection coefficient for the atom-field
system as follows \cite{14}
\begin{equation}
\begin{split}
r_{j}(\omega_{p}){=}\frac{[i(\omega_{c}{-}\omega_{p}){-}\frac{\kappa}{2}][i(\omega_{0}{-}\omega_{p}){+}\frac{\gamma}{2}]{+}g^{2}}
{[i(\omega_{c}{-}\omega_{p}){+}\frac{\kappa}{2}][i(\omega_{0}{-}\omega_{p}){+}\frac{\gamma}{2}]{+}g^{2}},
(j{=}L,R)
\end{split}
\end{equation}
where $\omega_{c}$ and $\omega_{p}$ are the frequencies of the
cavity and photon pulse, $\kappa$ and $\gamma$ are the cavity
damping rate and atomic decay rate respectively, and $g$ is the
atom-cavity coupling strength. Due to the large damping rate of
cavity, the absolute value of $r_{j}(\omega_{p})$ is verified to be
close to unity \cite{14}. This implies that the photon experiences a
very weak absorption, and thereby we may approximately consider that
the output photon only experiences a pure phase shift, i.e.,
$r_{j}(\omega_{p})=e^{i\phi}$, without any absorption.  On the other
hand, considering the case $g$=$0$ (the atom uncoupled to the cavity
or an empty cavity) we have
$r_{j0}(\omega_{p})=\frac{i(\omega_{c}{-}\omega_{p}){-}\frac{\kappa}{2}}
{i(\omega_{c}{-}\omega_{p}){+}\frac{\kappa}{2}}$ which can be
rewritten as a pure phase shift, i.e.,
$r_{j0}(\omega_{p}){=}e^{i\phi_{0}}$.

If the parameters of atom-field system satisfy
$\omega_{0}{=}\omega_{c}$, $\omega_{p}{=}\omega_{c}{-}\kappa/2$ and
$g{=}\kappa/2$, we can obtain $\phi{=}\pi$ and $\phi_{0}{=}\pi/2$,
corresponding to the evolution of atom and photon as
\begin{equation}
\begin{split}
&|L\rangle|g_{L}\rangle\rightarrow -|L\rangle|g_{L}\rangle, \  \
|R\rangle|g_{L}\rangle\rightarrow i|R\rangle|g_{L}\rangle,\\
&|L\rangle|g_{R}\rangle\rightarrow i|L\rangle|g_{R}\rangle, \ \ \
|R\rangle|g_{R}\rangle\rightarrow -|R\rangle|g_{R}\rangle.
\end{split}
\end{equation}
In the following ECPs, we will straightforwardly utilize the
evolution as shown in Eq. (2) without further illustration.
\begin{figure}[t]
\begin{center}
\scalebox{0.23}[0.23]{\includegraphics{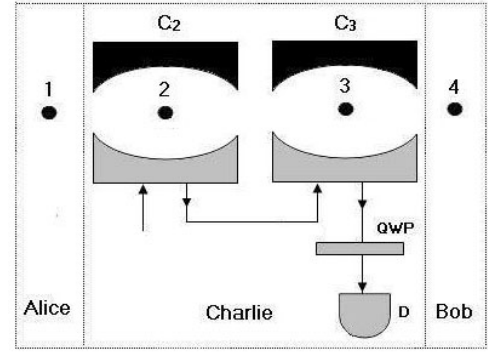}}
\end{center}
\caption{The schematic of atomic ECP. Atoms $2$ and $3$ are trapped
in low-Q cavities $C_{2}$ and $C_{3}$, respectively. A single-photon
pulse is sent into cavities $C_{2}$, $C_{3}$ and a quarter-wave
plate, then detected by photon detector $D$.}
\end{figure}

We now discuss concentration of atomic entangled states via photonic
Faraday rotation and the schematic setup is sketched in Fig. 2.
Assume that there are two pairs of non-maximally entangled
three-level atoms $1$, $2$ and $3$, $4$ as follows
\begin{equation}
\begin{split}
|\psi\rangle_{12}=a_{1}|g_{L}\rangle_{1}|g_{R}\rangle_{2}+b_{1}|g_{R}\rangle_{1}|g_{L}\rangle_{2},\\
|\psi\rangle_{34}=a_{2}|g_{L}\rangle_{3}|g_{R}\rangle_{4}+b_{2}|g_{R}\rangle_{3}|g_{L}\rangle_{4},
\end{split}
\end{equation}
where $a_{i}$ and $b_{i}$ $(i{=}1, 2)$ are the normalized
coefficients such that $|a_{i}|^{2}{+}|b_{i}|^{2}{=}1$, and we
assume that they are all real numbers without loss of generality. In
principle, the entangled states $|\psi\rangle_{12}$ and
$|\psi\rangle_{34}$, which are prepared by the same experimental
setup, have the identical amount of entanglement, i.e.,
$a_{1}{=}a_{2}$ and $b_{1}{=}b_{2}$. However, the experimental
imperfections or the effect of communication channels in the
preparation and distribution processes will lead to a tiny deviation
between $a_{1}(b_{1})$ and $a_{2}(b_{2})$. For simplicity, we
firstly omit the deviation and will discuss its effect to the
fidelity of our ECP later. We assume that three spatially separate
users, say Alice, Bob and Charlie, share entangled states
$|\psi\rangle_{12}$ and $|\psi\rangle_{34}$ where atoms $1$, $4$ are
in the hands of Alice and Bob respectively, and atoms $2$, $3$ are
all in the hand of Charlie.

To extract maximally entangled state from the pair of non-maximally
entangled states via photonic Faraday rotation, two low-Q cavities
$C_{2}$ and $C_{3}$, where atoms $2$ and $3$ are trapped
respectively, are introduced at Charlie's station. A single-photon
pulse with the initial state
$|\psi\rangle_{p}{=}\frac{1}{\sqrt{2}}(|L\rangle{+}|R\rangle)$ will
be sent through the cavities $C_{2}$ and $C_{3}$ sequentially. Then
Charlie performs the Hadamard operation on atoms 2, 3 and photon
respectively. Note that atomic Hadamard gate can be implemented by
driving the atom with an external classical field (polarized
lasers), and the quarter-wave plate (QWP) acts as the role of
photonic Hadamard gate. To be concrete, atomic and photonic Hadamard
operations can be expressed as $|g_{L}\rangle {\rightarrow}
\frac{1}{\sqrt{2}}(|g_{L}\rangle+|g_{R}\rangle)$, $|g_{R}\rangle
{\rightarrow} \frac{1}{\sqrt{2}}(|g_{L}\rangle{-}|g_{R}\rangle)$,
$|L\rangle{\rightarrow} \frac{1}{\sqrt{2}}(|L\rangle{+}|R\rangle)$
and $|R\rangle{\rightarrow}
\frac{1}{\sqrt{2}}(|L\rangle{-}|R\rangle)$. After this evolution
process, the quantum state of whole system is
\begin{equation}
\begin{split}
\sum_{j,k{=}L,R}\frac{1}{2}[{\mp}i|L\rangle|g_{j}\rangle_{2}|g_{k}\rangle_{3}
(a_{1}a_{2}|g_{L}\rangle_{1}|g_{R}\rangle_{4}{\pm}
b_{1}b_{2}|g_{R}\rangle_{1}|g_{L}\rangle_{4})&\\{+}
|R\rangle|g_{j}\rangle_{2}|g_{k}\rangle_{3}(\mp
a_{1}b_{2}|g_{L}\rangle_{1}|g_{L}\rangle_{4}{+}
b_{1}a_{2}|g_{R}\rangle_{1}|g_{R}\rangle_{4}).&
\end{split}
\end{equation}
Finally, Charlie performs measurement on the states of photon and
atoms at his side, and thus the atomic state at Alice's and Bob's
sides will collapse into one of the corresponding components in Eq.
(4). To be explicit, if Charlie's measurement outcome is
$|L\rangle|g_{j}\rangle_{2}|g_{k}\rangle_{3}$ $(j, k{=}L, R)$, the
quantum state of atoms $1$, $4$ will be $|\psi\rangle_{14}{=}
a_{1}a_{2}|g_{L}\rangle_{1}|g_{R}\rangle_{4}{\pm}
b_{1}b_{2}|g_{R}\rangle_{1}|g_{L}\rangle_{4}$ (unnormalized). On the
other hand, if Charlie's measurement outcome is
$|R\rangle|g_{j}\rangle_{2}|g_{k}\rangle_{3}$ $(j, k{=}L, R)$, the
quantum state of atoms $1$, $4$ will be $|\psi'\rangle_{14}{=}\mp
a_{1}b_{2}|g_{L}\rangle_{1}|g_{L}\rangle_{4}+
b_{1}a_{2}|g_{R}\rangle_{1}|g_{R}\rangle_{4}$. Obviously,
$|\psi'\rangle_{14}$ are maximally entangled under the previous
condition of the initial states, and thus the total successful
probability of our ECP is $P{=}2a_{1}^{2}(1{-}a_{1}^{2})$, which is
the same as that in Ref. \cite{9}.

In this atomic ECP, two atoms $1$, $4$, which never interacted with
each other before, are left in a pure maximally entangled state
after the whole operation process. In other words, Alice and Bob are
completely passive in the whole concentration process. Starting from
this point of view, we can generalize this ECP to reconstruct
multi-atom Greenberger-Horne-Zeilinger (GHZ) state from the
partially entangled atomic GHZ-class states as follows
\begin{equation}
\begin{split}
|\Psi\rangle_{1}=a_{1}|g_{L}g_{R}{\cdots}g_{R}\rangle_{C_{1}A_{1}{\cdots}A_{N}}{+}b_{1}|g_{R}g_{L}{\cdots}g_{L}\rangle_{C_{1}A_{1}{\cdots}A_{N}},\\
|\Psi\rangle_{2}=a_{2}|g_{L}g_{R}{\cdots}g_{R}\rangle_{C_{2}B_{1}{\cdots}B_{N}}{+}b_{2}|g_{R}g_{L}{\cdots}g_{L}\rangle_{C_{2}B_{1}{\cdots}B_{N}},
\end{split}
\end{equation}
where the subscripts $A_{j}$, $B_{j}$$(j{=}1,...,N)$, $C_{1}$ and
$C_{2}$ represent atoms held by Alice, Bob and Charlie,
respectively. If we define
$|g_{L}'\rangle_{A}{=}|g_{L}{\cdots}g_{L}\rangle_{A_{1}{\cdots}
A_{N}}$,
$|g_{R}'\rangle_{A}{=}|g_{R}{\cdots}g_{R}\rangle_{A_{1}{\cdots}
A_{N}}$,
$|g_{L}'\rangle_{B}{=}|g_{L}{\cdots}g_{L}\rangle_{B_{1}{\cdots}
B_{N}}$ and
$|g_{R}'\rangle_{B}{=}|g_{R}{\cdots}g_{R}\rangle_{B_{1}{\cdots}
B_{N}}$, the GHZ-class states can be rewritten as
$|\Psi\rangle_{1}=a_{1}|g_{L}\rangle_{C_{1}}|g_{R}'\rangle_{A}{+}b_{1}|g_{R}\rangle_{C_{1}}|g_{L}'\rangle_{A}$
and
$|\Psi\rangle_{2}=a_{2}|g_{L}\rangle_{C_{2}}|g_{R}'\rangle_{B}{+}b_{2}|g_{R}\rangle_{C_{2}}|g_{L}'\rangle_{B}$.
By inspection, they just have the same forms as the entangled states
in Eq. (3). Thus, we can adopt the same procedure as that in the
case of two-atom entangled state, and reconstruct $2N$-atom GHZ
state between Alice and Bob with the successful probability
$P{=}2a_{1}^{2}(1{-}a_{1}^{2})$.

\begin{figure}[t]
\begin{center}
\scalebox{0.24}[0.24]{\includegraphics{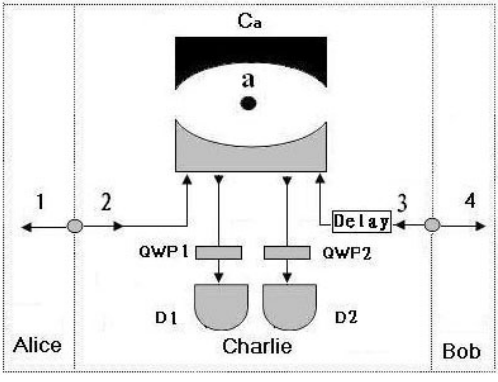}}
\end{center}
\caption{The schematic of photonic ECP. Atoms $a$ trapped in low-Q
cavity $C_{a}$, Quarter-Wave plate QWP1, QWP2, and photon detectors
$D_{1}$, $D_{2}$ are introduced at Charlie's station.}
\end{figure}

Benefitting from the single-photon input-output process in the cavity QED system, we show that photonic Faraday rotation can also be used
to concentrate photonic entangled states. The concrete schematic setup for photonic ECP is depicted in Fig. 3. We assume that there are
two pairs of partially entangled photonic states as follows
\begin{equation}
\begin{split}
|\phi\rangle_{12}{=}a_{1}|L\rangle_{1}|R\rangle_{2}{+}b_{1}|R\rangle_{1}|L\rangle_{2},\\
|\phi\rangle_{34}{=}a_{2}|L\rangle_{3}|R\rangle_{4}{+}b_{2}|R\rangle_{3}|L\rangle_{4},
\end{split}
\end{equation}
where photons $1$, $4$ are in the hands of Alice and Bob
respectively, and Charlie holds photons $2$ and $3$. To implement
photonic ECP, a three-level atom (labeled as $a$), which is trapped
in a low-Q cavity $C_{a}$, is introduced at Charlie's station. The
initial state of atom is
$|\phi\rangle_{a}{=}\frac{1}{\sqrt{2}}(|g_{L}\rangle_{a}{+}|g_{R}\rangle_{a})$.
Charlie guides photons $2$, $3$ into the cavity sequentially, and
then lets them pass QWP1 and QWP2 respectively after leaving the
cavity $C_{a}$. By performing the Hadamard operation on atom $a$,
the quantum state after this evolution process will be
\begin{equation}
\begin{split}
\sum_{j,k{=}L,R}\frac{1}{2}[{\mp}i|j\rangle_{2}|k\rangle_{3}|g_{L}\rangle_{a}(a_{1}a_{2}|L\rangle_{1}|R\rangle_{4}{+}b_{1}b_{2}|R\rangle_{1}|L\rangle_{4})\\
{+}|j\rangle_{2}|k\rangle_{3}|g_{R}\rangle_{a}(a_{1}b_{2}|R\rangle_{1}|R\rangle_{4}{\mp}b_{1}a_{2}|L\rangle_{1}|L\rangle_{4})].
\end{split}
\end{equation}
Charlie then measures the quantum states of photons and atom at his
side, followed by the collapse of photonic states at Alice's and
Bob's side to one of the corresponding components in Eq. (7). In
detail, the quantum state of photons $1$, $4$ will be
$|\psi\rangle_{14}{=}a_{1}a_{2}|L\rangle_{1}|R\rangle_{4}{+}b_{1}b_{2}|R\rangle_{1}|L\rangle_{4}$
or
$|\psi'\rangle_{14}{=}a_{1}b_{2}|R\rangle_{1}|R\rangle_{4}{\mp}b_{1}a_{2}|L\rangle_{1}|L\rangle_{4
}$ corresponding to the measurement outcomes
$|j\rangle_{2}|k\rangle_{3}|g_{L}\rangle_{a}$ or
$|j\rangle_{2}|k\rangle_{3}|g_{R}\rangle_{a}$ $(j, k{=}L, R)$,
respectively. Similar to atomic ECP, $|\psi'\rangle_{14}$ are
maximally entangled and then the successful probability of our
photonic ECP is $P{=}2a_{1}^{2}(1{-}a_{1}^{2})$.

In this photonic ECP, Charlie needs to strictly control the time
interval of photons $2$ and $3$ passing the low-Q cavity, in order
to avoid the case that both of photons interact with the atom
simultaneously. Otherwise, the ECP will fail. However, the order of
photon ($2$ or $3$) interacting with atom $a$ will not affect the
final results of ECP because the situation of photon $2$ and $3$ is
completely equivalent. Similar to the atomic ECP, we can also
generalize the photonic ECP to reconstruct multi-photon GHZ state
from the partially entangled photonic GHZ-class states
$|\Phi\rangle_{1}{=}a_{1}|LR{\cdots}R\rangle_{C_{1}A_{1}{\cdots}A_{N}}{+}b_{1}|RL{\cdots}L\rangle_{C_{1}A_{1}{\cdots}A_{N}}$
and
$|\Phi\rangle_{2}{=}a_{2}|LR{\cdots}R\rangle_{C_{2}B_{1}{\cdots}B_{N}}{+}b_{2}|RL{\cdots}L\rangle_{C_{2}B_{1}{\cdots}B_{N}}$
with the same successful probability. It is noted that our ECPs for
atomic and photonic states may be universal as the entanglement can
be concentrated whenever $a_i{<}b_i$ or $a_i{>}b_i$ for every
$i{=}1,2$.

We briefly discuss the experimental feasibility of our protocols.
Consider a $^{87}$Rb atom trapped in the fiber-based Fabry-Perot
cavity \cite{21}. The states
$|\mathrm{F}{=}2,m_{\mathrm{F}}{=}{\pm}1\rangle$ of level $5S_{1/2}$
correspond to degenerate ground states $|g_{L}\rangle$ and
$|g_{R}\rangle$ respectively, the state
$|\mathrm{F}'{=}3,m_{\mathrm{F}}{=}0\rangle$ of level $5P_{3/2}$ is
chosen as the excited state $|e\rangle$ and the corresponding
transition frequency $\omega_{0}{=}2\pi c/\lambda$ with
$\lambda{=}780$nm (D$_2$ line). In Ref. \cite{21}, the cavity length
$L{=}38.6\mu$m, waist radius $w_{0}{=}3.9\mu$m and finesse
$\mathcal{F}{=}37000$, which correspond to longitudinal mode number
$n{=}99$, the cavity decay rate $\kappa{=}2\pi{\times}53$MHz (the
relevant Q factor $Q{=}\omega_c/(2\kappa){=}3.63{\times}10^6$) and
the maximal coupling strength $g_{0}{=}2\pi{\times}215$MHz. In our
protocols, the atom-cavity coupling strength $g{=}g_{0}\cos(2\pi
x/\lambda)$ should be matched with cavity decay rate
($g{=}\kappa/2$), which can be satisfied by adjusting the
appropriate atomic longitudinal coordinate
($x{=}n\frac{\lambda}{2}{+}179$nm). Meanwhile, the input photon can
be tuned to be nearly resonant with the atom-cavity system, i.e.,
$\omega_{p}{=}\omega_{c}{-}\kappa/2$. Therefore, based on present
experiment technology in cavity QED \cite{13,21}, the required
atom-cavity parameters can be tuned to control the reflectivity of
the input photon for obtaining the desired phase shifts. In the
following, we consider the possible realization of our ECPs in the
context of low-Q cavity. In the experiment, the cavity Q factor and
the decay rate are closely related with the cavity finesse which
depends solely on the intensity transmission and loss of cavity
mirror. In Ref. \cite{21}, if the atom is located at the antinode of
the cavity field ($x{=}n\frac{\lambda}{2}$), we can obtain the
maximal atom-cavity coupling ($g{=}g_{0}{=}2\pi{\times}215$MHz).
Consider the transmission of cavity mirror $\mathcal {T}{=}666$ppm
(i.e., the cavity finesse $\mathcal{F}{=}4510$), the practical Q
factor of cavity reduces to only $Q{=}4.47\times10^5$ and then the
decay rate satisfies $\kappa{=}2g_{0}$. As to the ultra low-Q cavity
(high decay rate of cavity $\kappa$), the condition $g{=}\kappa/2$
may also be satisfied by obtaining the large enough coupling between
atom and cavity field.

However, there are still some imperfections in the realistic
experiment. For instance, the cavity resonance frequency may be
deviated from the atomic eigenfrequency due to the tiny change of
cavity length, and the coupling strength may be not strictly matched
with the cavity decay rate because of the variation of atomic
position in the cavity. The slight deviation of resonance
($\omega_{c}{\sim}\omega_{0}$) and mismatch of coupling strength
($g{\sim}\kappa/2$) will not change the reflection amplitudes but
phase shifts $\phi(\phi_{0})$. In the case of
$\omega_{c}{-}\omega_{0}{\approx}\kappa/10$, the phase shifts
$\phi{\approx}2.75$ and $\phi_{0}{\approx}1.36$. The fidelity of
obtaining atomic and photonic states $|\psi\rangle_{14}$ is about
$F{=}\frac{1}{2}[1{-\cos2(\phi{-}\phi_{0})}]{\approx}0.955$. If the
coupling strength satisfies $g{\approx}3\kappa/5$, we can obtain the
phase shift $\phi{\approx}2.31$ and the fidelity of
$|\psi\rangle_{14}$ $F{\approx}0.455$. Interestingly, the fidelity
of quantum state $|\psi'\rangle_{14}$ is just $1$ as it is
independent of the Faraday rotation angle. Thus, our atomic and
photonic ECPs are immune to the experimental imperfections as
discussed above.

In our atomic and photonic ECPs, we have assumed that the initial
condition of the entangled states satisfies $a_{1}{=}a_{2}$ and
$b_{1}{=}b_{2}$. But in practice, there may be imperfections in the
entanglement preparation and distribution processes, which lead the
entangled states into less entangled pure or even mixed  ones. Here,
we consider that the entanglement preparation process is near
perfect and the communication channels between Alice (Bob) and
Charlie are of high quality. Then the resulting entangled states,
after the entanglement preparation and distribution processes, may
have a tiny deviation to the ideal ones, i.e.,
$a_{2}{=}a_{1}{+}ka_{1}$ with $k$ being a small constant. In this
case, the fidelity of obtaining the desired state
$|\psi'\rangle_{14}$ is
$F(a_{1},k){=}\frac{[\sqrt{1{-}a^{2}_{1}(1{+}k)^{2}}{+}(1{+}k)\sqrt{1{-}a^{2}_{1}}]^2}{2[1{+}(1{+}k)^{2}{-}2a^{2}_{1}(1{+}k)^{2}]}$
\cite{9}. If we consider $a_{1}\in(0,0.7)$ and $k{=\pm}0.1$, the
minimal fidelity $F{=}0.989,0.991$ for $a_{1}{=}0.7$ and
$k{=\pm}0.1$, which indicates that the small deviation of
coefficients, due to the effect of imperfections described above,
only affects the fidelity of the result state slightly.

In the following, we make comparison with the previous ECPs. Note
that ECPs involving a pair of partially entangled states can be
realized via entanglement swapping \cite{20}. The crucial step of
entanglement swapping is the implementation of joint Bell-state
measurement, which is also at the heart of other QIP tasks such as
quantum teleportaion \cite{1} and dense coding \cite{3}. In our
atomic and photonic ECPs, we have introduced low-Q cavities,
three-level atom and single-photon pulse, and can implement
entanglement swapping without joint Bell-state measurement, only by
detecting the quantum state of atoms and photons separately.

For concentration of atomic entanglement, the distinct advantage of
our atomic ECP is that we only need low-Q optical cavity while the
high-Q cavity is usually required in Refs. \cite{9,10,11}. In Refs.
\cite{9,11}, the atomic state is used as the flying qubit, but it is
actually suitable for acting as stationary qubit which will be
feasible in experiment. In Ref. \cite{10}, Cao \emph{et al.}
proposed atomic ECP through cavity decay which relies on two leaking
photon reaching the beam splitter simultaneously, as well as the
high efficiency of two photon detectors. Due to the large
inefficiency of photon detector, our atomic ECP may be more
efficient than Ref. \cite{10} as only a single-photon detector is
involved for ours. Furthermore, we can use the coherent input pulse
to replace the single-photon pulse as shown in Ref. \cite{17}, and
also implement atomic ECP with homodyne detection of coherent light,
which can greatly relaxes the experiment requirement for photon
source and reduce measurement difficulties. On the other hand, we
propose to concentrate photonic entanglement via single-photon
input-output process in cavity QED for the first time. With the
assistance of three-level atom trapped in the low-Q cavity,
two-photon and multi-photon maximally entangled states can be
reconstructed with the same efficiency (successful probability) as
that in Ref. \cite{6}. In fact, the low-Q cavity and three-level
atom function as photonic phase-shift controller in photonic Faraday
rotation, which is similar to the cross-kerr nonlinearity \cite{23}.
Therefore, we can also construct photonic parity gate via photonic
Faraday rotation and then implement photonic ECP as Ref. \cite{8}.

In conclusion, we have proposed to concentrate atomic and photonic
entanglement via photonic Faraday rotation. Through the
single-photon input-output process in cavity QED, it is shown that
the maximally entangled atomic and photonic state can be extracted
from two partially entangled states. In our ECPs, we only need the
low-Q cavity, three-level atom and the basic optical elements such
as QWP and photon detector to complete entanglement concentration,
and they may be feasible with current cavity QED and quantum optics
technology. The essential idea in our ECPs is the single-photon
input-output process in cavity QED, and thus it may be worth
studying entanglement purification and concentration using other
similar cavity QED schemes \cite{24} in the future.

This work was partially supported by the National Fundamental
Research Program of China (Grants No. 2007CB925204), NSFC (Grants
Nos. 11075050 and 10974016), the Program for Changjiang Scholars and
Innovative Research Team in University under Grant No. IRT0964, and
Hunan Provincial Natural Science Foundation under Grant No.
11JJ7001, and the Scientific Research Fund of Hunan Provincial
Education Department, China (Grants Nos. 10A032 and 10C0698).

\end{document}